\documentclass[aps,prd,twocolumn,preprintnumbers,amssymb,nobibnotes,nofootinbib,longbibliography,superscriptaddress]{revtex4-2}
\usepackage[utf8]{inputenc}
\usepackage[T1]{fontenc}

\usepackage{amsmath, amsfonts, amssymb}
\usepackage{mathtools}
\usepackage{physics}
\usepackage{graphicx}
\usepackage{svg}
\usepackage{booktabs}
\usepackage{array}
\usepackage{enumitem}
\usepackage{hyperref}

\hypersetup{%
    colorlinks=true,%
    allcolors={blue!60!black}%
}

\newcommand{\as}{a_s}
\newcommand{\CS}{\textcolor{blue!60!black}}
\newcommand{\nf}{n_f}
\newcommand{\ns}{\mathrm{ns}}
\newcommand{\ps}{\mathrm{ps}}

\allowdisplaybreaks
\usepackage{cleveref}

\graphicspath{ {figures/} }
\newcommand{\FDiag}[2]{
\begin{minipage}{0.155\textwidth}
\begin{center}
\CS{#1}\\[-2ex]
\includegraphics[angle=-90,width=\textwidth]{#2}
\end{center}
\end{minipage}
\hspace*{-2ex}
}

\begin{document}

\title{The four-loop quark-to-quark splitting function in QCD}

\author{Thomas~Gehrmann}
\email{thomas.gehrmann@uzh.ch}
\affiliation{Physik-Institut, Universit\"at Z\"urich, Winterthurerstrasse 190, 8057 Z\"urich, Switzerland}

\author{Andreas von Manteuffel}
\email{manteuffel@ur.de}
\affiliation{Institut f\"ur Theoretische Physik, Universit\"at Regensburg, 93040 Regensburg, Germany}

\author{Vasily Sotnikov}
\email{sotnikov@uni-mainz.de}
\affiliation{Institute of Physics, Johannes Gutenberg University Mainz, Staudinger Weg 7, 55099 Mainz, Germany}
\affiliation{Physik-Institut, Universit\"at Z\"urich, Winterthurerstrasse 190, 8057 Z\"urich, Switzerland}

\author{Tong-Zhi Yang}
\email{tongzhi.yang@m.scnu.edu.cn}
\affiliation{State Key Laboratory of Nuclear Physics and Technology, Institute of Quantum Matter,
South China Normal University, Guangzhou 510006, China}

\date{\today}

\begin{abstract}
The scale evolution of parton distributions is governed by splitting functions.
We compute the four-loop pure-singlet contribution to the quark-to-quark splitting function in perturbative QCD.
Together with the known non-singlet contribution, our result completes the quark-to-quark splitting function at this order.
We confirm previous partial results and obtain, for the first time, a fully analytic expression valid for all momentum fractions.
We examine its behavior at small momentum fractions and provide precise numerical representations suitable for parton evolution.
\end{abstract}

\maketitle

\preprint{ZU-TH 37/26}
\preprint{MITP-26-049}

\section{Introduction}

Parton distribution functions (PDFs) encode the quark and gluon content of hadrons and are essential inputs to predictions for hadron collisions.
Their uncertainties limit the precision of these predictions, motivating improvements in both their determination from data and their perturbative evolution.
The dependence of PDFs on the factorization scale is governed by the QCD splitting functions through the DGLAP equations~\cite{Altarelli:1977zs,Gribov:1972ri,Dokshitzer:1977sg}, with four-loop splitting functions required for consistent evolution at third order in QCD perturbation theory (next-to-next-to-next-to leading order, N$^3$LO).
The difficulty of their calculation has motivated approximations based on fixed Mellin moments, partial analytic results, and the known behavior at small and large momentum fractions~\cite{Catani:1994sq,Moch:2017uml,Moch:2021qrk,Falcioni:2023luc,Falcioni:2023vqq,Moch:2023tdj,Falcioni:2024xyt,Falcioni:2024qpd,Falcioni:2025hfz}.
These approximations leave a residual theoretical uncertainty in PDF evolution, particularly prominent at small momentum fractions.
At four loops, complete analytic results have so far been obtained only for the non-singlet quark distributions, which evolve without mixing with the gluon distribution~\cite{Gehrmann:2026qbl}.
In this work, we determine the complete pure-singlet contribution analytically, thereby completing the four-loop quark-to-quark splitting function and eliminating the approximation uncertainty in this entry of the evolution matrix.

We work in the $\overline{\mathrm{MS}}$ scheme with $\nf$ massless quark flavors and identify the factorization and renormalization scales, denoted by $\mu$.
The quark singlet distribution $q_s=\sum_{i=1}^{\nf}(q_i+\bar q_i)$ and the gluon distribution $g$ obey the coupled evolution equations
\begin{equation}
 \frac{\dd}{\dd\log\mu^2}
 \begin{pmatrix}q_s\\g\end{pmatrix}
 =
 \begin{pmatrix}P_{qq}&P_{qg}\\P_{gq}&P_{gg}\end{pmatrix}
 \otimes
 \begin{pmatrix}q_s\\g\end{pmatrix},
 \label{eq:dglap}
\end{equation}
where $\otimes$ denotes Mellin convolution in the momentum fraction $x$.
We expand the splitting functions as $P_{ij}(x)=\sum_{k\geq0}\as^{k+1}P_{ij}^{(k)}(x)$, with $\as=\alpha_s(\mu^2)/(4\pi)$, and define the corresponding anomalous dimensions by
\begin{equation}
 \gamma_{ij}^{(k)}(n)
 =-\int_0^1\dd x\,x^{n-1}P_{ij}^{(k)}(x).
 \label{eq:conventions}
\end{equation}
The quark-to-quark splitting function decomposes as $P_{qq}=P_{\ns}^{+}+P_{\ps}$.
Here $P_{\ns}^{+}$ governs the evolution of the charge-conjugation-even flavor non-singlet combinations $q_i+\bar q_i-q_j-\bar q_j$, while the pure-singlet contribution $P_{\ps}$ first appears at two loops.

Several partial results for $P_{\ps}^{(3)}$ preceded the present calculation.
The contributions proportional to $\nf^3$ and $\nf^2$ were determined analytically for general $n$ in Refs.~\cite{Davies:2016jie} and~\cite{Gehrmann:2023cqm}, respectively.
Fixed-moment calculations provided the quartic-color contribution through $n=16$~\cite{Moch:2018wjh} and the four lowest even moments of the full singlet evolution matrix~\cite{Moch:2021qrk}.
The pure-singlet results were subsequently extended to all even moments through $n=20$~\cite{Falcioni:2023luc} and then to $n=22$~\cite{Falcioni:2025hfz}.
Analytic expressions for the coefficients of the explicit $\zeta_3$, $\zeta_4$, and $\zeta_5$ terms in $\gamma_{\ps}^{(3)}$ are also available for general $n$~\cite{Falcioni:2023luc,Falcioni:2025hfz}.
Further information on the small-$x$ behavior comes from high-energy and double-logarithmic resummation~\cite{Catani:1994sq,Bonvini:2018xvt,Davies:2022ofz}.
These results provide independent checks of our calculation.

\section{Methodology}

\begin{figure*}[]
\begin{center}
 \FDiag{$n_f C_F C_A^2$}{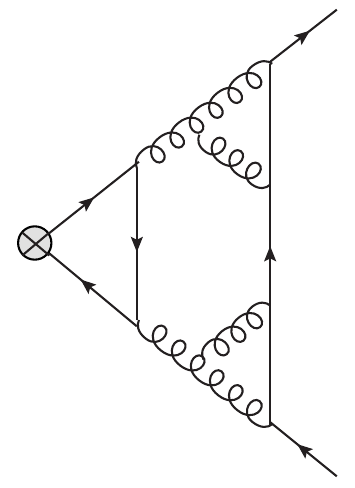}
 \FDiag{$n_f C_F^2 C_A$}{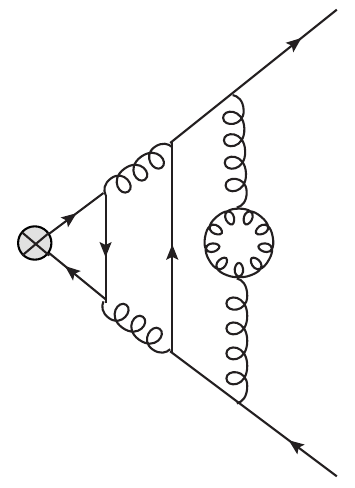}
 \FDiag{$n_f C_F^3$}{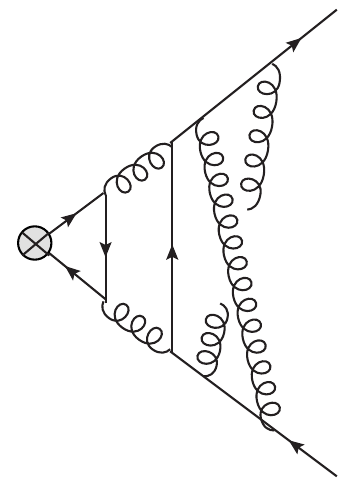}
 \FDiag{$\nf (d_F^{abcd})^2/N_c$}{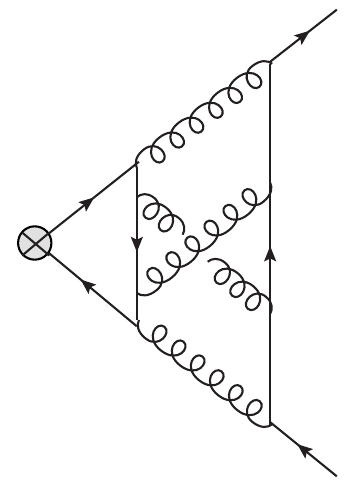}
\\[1ex]
 \FDiag{$\nf^2 C_F C_A$}{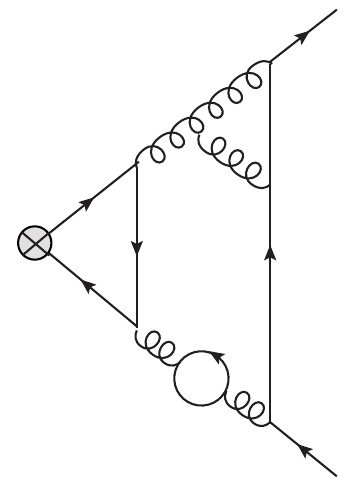}
 \FDiag{$\nf^2 C_F^2$}{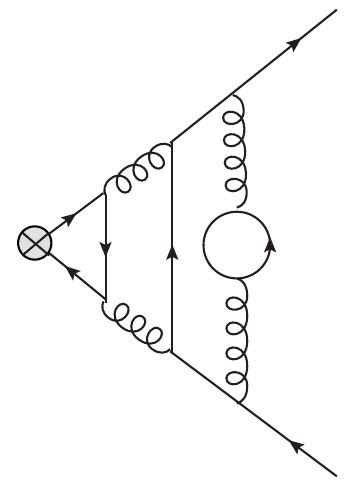}
 \FDiag{$\nf^3 C_F$}{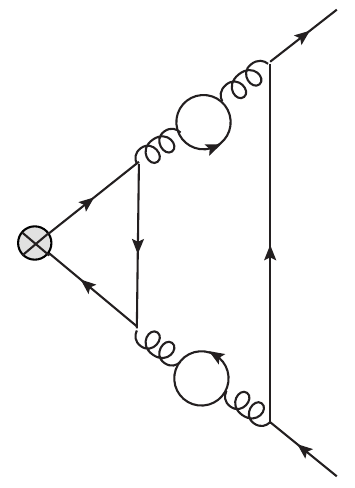}
\end{center}
\caption{\label{fig:diags}Sample Feynman diagrams for off-shell quark self energies with an operator insertion (crossed vertex) contributing to different color coefficients of the splitting functions $P_\ps^{(3)}$ at four loops.
Both planar and non-planar diagrams contribute.
In further diagrams not shown here, the operator insertion may also involve gluons (curly lines) in addition to the two quarks (straight lines).
}
\end{figure*}

The splitting function $P_{qq}(x)$ is related by a Mellin transform to the anomalous dimension $\gamma_{qq}(n)$ of the flavor-singlet twist-two quark operator
\begin{equation}
\label{eq:quarkOP}
O_q(n) = \frac{i^{n-1}}{2}\,
\bar{\psi}\,\Delta\cdot\gamma\,(\Delta\cdot D)^{n-1}\psi\,,
\end{equation}
where $D$ is the covariant derivative, $\Delta$ is an auxiliary lightlike vector, and the sum over quark flavors is implicit.
Under renormalization, $O_q$ mixes with the physical gluon operator $O_g$ and with gauge-variant operators~\cite{Dixon:1974ss,Kluberg-Stern:1974iel,Joglekar:1975nu,Collins:1994ee,Falcioni:2022fdm,Gehrmann:2023ksf}.
Its renormalization takes the form~\cite{Gehrmann:2023ksf}
\begin{equation}
\label{eq:RenorGVSimQ}
O_q^{\mathrm{R}}
= Z_{qq} O_q^{\mathrm{B}}
+ Z_{qg} O_g^{\mathrm{B}}
+ Z_{qA} O_{ABC}^{\mathrm{B}}
+ [ZO]_q^{\mathrm{GV}}\,,
\end{equation}
where the superscripts '$\mathrm{R}$' and '$\mathrm{B}$' denote renormalized and bare operators.
The constants $Z_{qq}$ and $Z_{qg}$ describe mixing among physical operators, while $Z_{qA}$ describes mixing with the combination $O_{ABC}$ of gauge-variant gluon, quark, and ghost operators.
The term $[ZO]_q^{\mathrm{GV}}$ collects additional gauge-variant counterterms.
Its two-quark matrix element vanishes through four loops, so it can be omitted when extracting $\gamma_{qq}^{(3)}$~\cite{Gehrmann:2023ksf}.

The physical anomalous dimensions are defined by
\begin{equation}
\label{eq:anomdimdef}
\frac{\dd Z(\mu,n)}{\dd\log\mu^2}
= -\gamma(\mu,n) \cdot \,Z(\mu,n) \,,
\end{equation}
where $Z$ and  $\gamma$ are matrices with indices running over $\{q,g\}$.
Within the operator product expansion, the anomalous dimensions are extracted from the ultraviolet poles of the corresponding off-shell two-parton operator matrix elements (OMEs)~\cite{Gross:1974cs,Politzer:1974fr}.
The explicit expression for $Z$ in terms of the anomalous dimensions reads 
\begin{align}
Z={}&1+\as\,\frac{\gamma^{(0)}}{\epsilon}
+\frac{\as^2}{2}\bigg\{
 \frac{\gamma^{(1)}}{\epsilon}
 +\frac{(\gamma^{(0)})^2-\beta_0\gamma^{(0)}}{\epsilon^2}
\bigg\}
\nonumber\\
&+\frac{\as^3}{6}\bigg\{
 \frac{2\gamma^{(2)}}{\epsilon}
 +\frac{1}{\epsilon^2}\Big[
 2\gamma^{(1)}\gamma^{(0)}+\gamma^{(0)}\gamma^{(1)}
\nonumber\\
&\qquad-2\beta_0\gamma^{(1)}-2\beta_1\gamma^{(0)}
 \Big]
\nonumber\\
&\quad+\frac{1}{\epsilon^3}\Big[
 (\gamma^{(0)})^3-3\beta_0(\gamma^{(0)})^2
 +2\beta_0^2\gamma^{(0)}
 \Big]\bigg\}
\nonumber\\
&+\frac{\as^4}{24}\bigg\{
 \frac{6\gamma^{(3)}}{\epsilon}
 +\frac{1}{\epsilon^2}\Big[
 6\gamma^{(2)}\gamma^{(0)}+2\gamma^{(0)}\gamma^{(2)}
\nonumber\\
&\qquad+3(\gamma^{(1)})^2-6\beta_0\gamma^{(2)}
 -6\beta_1\gamma^{(1)}-6\beta_2\gamma^{(0)}
 \Big]
\nonumber\\
&\quad+\frac{1}{\epsilon^3}\Big[
 3\gamma^{(1)}(\gamma^{(0)})^2
 +2\gamma^{(0)}\gamma^{(1)}\gamma^{(0)}
 +(\gamma^{(0)})^2\gamma^{(1)}
\nonumber\\
&\qquad-9\beta_0\gamma^{(1)}\gamma^{(0)}
 -5\beta_0\gamma^{(0)}\gamma^{(1)}
 -8\beta_1(\gamma^{(0)})^2
\nonumber\\
&\qquad+6\beta_0^2\gamma^{(1)}
 +12\beta_0\beta_1\gamma^{(0)}
 \Big]
\nonumber\\
&\quad+\frac{1}{\epsilon^4}\Big[
 (\gamma^{(0)})^4-6\beta_0(\gamma^{(0)})^3
 +11\beta_0^2(\gamma^{(0)})^2
\nonumber\\
&\qquad-6\beta_0^3\gamma^{(0)}
 \Big]\bigg\}
+\mathcal{O}(\as^5)\,.
\label{eq:ZfactorIntermsofGamma}
\end{align}

We therefore determine $\gamma_{qq}^{(3)}$ by extracting $Z_{qq}$ from the renormalization of the two-quark OME, which through four loops reads
\begin{multline}
\label{eq:2qomeForOqRenor}
\mel{q}{O_q^{\mathrm{R}}}{q}
= Z_q \bigg[
Z_{qq}\mel{q}{O_q^{\mathrm{B}}}{q}
\\
+ Z_{qg}\mel{q}{O_g^{\mathrm{B}}}{q}
+ Z_{qA}\mel{q}{O_{ABC}^{\mathrm{B}}}{q}
\bigg]\Bigg\vert_{a_s^{\mathrm{B}}\to Z_{a_s}a_s}\,.
\end{multline}
Here $Z_q$ and $Z_{a_s}$ are the quark wave-function and strong-coupling renormalization constants, respectively.
The gauge-parameter renormalization, governed by the gluon wave-function renormalization constant $Z_g$, is left implicit.
Explicit expressions for these constants are collected in Appendix~A of Ref.~\cite{Gehrmann:2023ksf}.

The only missing ingredient for determining $Z_{qq}$ at four loops is the four-loop OME $\mel{q}{O_q^{\mathrm{B}}}{q}$, which we compute in this work.
All other contributions in \cref{eq:2qomeForOqRenor} are known to the required orders.
We use $\mel{q}{O_g^{\mathrm{B}}}{q}$ through three loops from Ref.~\cite{Gehrmann:2023ksf}.
Since $Z_{qA}$ starts at order $a_s^2$ and $\mel{q}{O_{ABC}^{\mathrm{B}}}{q}$ starts at order $a_s$, the gauge-variant contribution requires $\mel{q}{O_{ABC}^{\mathrm{B}}}{q}$ through two loops and $Z_{qA}$ through three loops, both available in Ref.~\cite{Gehrmann:2023ksf}.
Working in dimensional regularization with $d=4-2\epsilon$, we extract $\gamma_{qq}^{(3)}$ from the coefficient of the $1/\epsilon$ pole in the four-loop contribution to $Z_{qq}$.

We compute the off-shell OME using the strategy of Ref.~\cite{Gehrmann:2026qbl}.
We generate the contributing Feynman diagrams with \texttt{Qgraf} \cite{Nogueira:1991ex}.
Example diagrams are shown in \cref{fig:diags}.
The Lorentz and Dirac algebra is performed with \texttt{Form}~\cite{Vermaseren:2000nd,Davies:2026cci}, and the color algebra with \texttt{Color.h}~\cite{vanRitbergen:1998pn}.
An auxiliary parameter $t$ converts the symbolic powers, such as $(\Delta\cdot k)^{n-1}$ generated by a spin-$n$ operator into linear propagators. 
The individual Mellin moments are subsequently obtained from the expansion in $t$~\cite{Ablinger:2012qm,Ablinger:2014nga}.
We organize the resulting scalar integrals into families and sectors, identify integrals related by momentum shifts, and perform integration-by-parts reductions~\cite{Chetyrkin:1980pr,Laporta:2000dsw} with \texttt{Reduze 2}~\cite{vonManteuffel:2012np} and \texttt{Finred}.
The matrix-element reduction uses finite-field sampling~\cite{vonManteuffel:2014ixa,Peraro:2016wsq} and optimized selections of reduction equations~\cite{Driesse:2024xad,Guan:2024byi,Bern:2024adl,vonHippel:2025okr,Song:2025pwy,Lange:2025fba}.
For each top-level sector we sample a set of spanning sectors~\cite{Larsen:2015ped,Guan:2024byi}.
The number of top sectors and spanning sectors required here is of the same order of magnitude as encountered in the non-singlet anomalous dimension \cite{Gehrmann:2026qbl},
with a marginally smaller number of required master integrals.
We choose basis integrals to limit denominators with mixed $\epsilon$ and $t$ dependence~\cite{Smirnov:2020quc,Usovitsch:2020jrk}, then reconstruct the reduction coefficients after determining their denominators and simple numerator factors~\cite{Abreu:2018zmy,Heller:2021qkz}.
We derive differential equations in $t$ for the basis integrals~\cite{Gehrmann:1999as,Henn:2013pwa} and solve them as Laurent series in $\epsilon$ whose coefficients are Taylor series in $t$.
Recursion relations obtained from the differential equations determine the Taylor coefficients~\cite{Blumlein:2009tj,Ablinger:2015tua,Lee:2017qql}.
At $t=0$, the required regular solutions reduce to standard four-loop self-energy integrals, which fix the boundary conditions~\cite{Baikov:2010hf,Lee:2011jt,vonManteuffel:2019gpr,Lee:2023dtc}.

We use the recursion relations to generate sufficiently many powers of $t$ to determine the bare four-loop quark-singlet OMEs at high Mellin moments.
We extract $\gamma_{qq}^{(3)}$ from their poles after accounting for quark--gluon operator mixing, gauge-variant counterterms, and the required lower-order matrix elements~\cite{Gehrmann:2023ksf}.
Subtracting the known flavor non-singlet contribution~\cite{Gehrmann:2026qbl} gives $\gamma_{\ps}^{(3)}=\gamma_{qq}^{(3)}-\gamma_{\ns}^{+,(3)}$.
We reconstruct its all-$n$ dependence using a generic ansatz of harmonic sums~\cite{Vermaseren:1998uu} up to weight six and rational prefactors that allow denominator factors involving $n$, $n\pm 1$, and $n\pm 2$.

\section{Results}

We successfully infer all coefficients in the 
ansatz described above from the calculated Mellin moment samples and confirm that harmonic sums \cite{Vermaseren:1998uu} up to weight six appear.

Our results are consistent with the fixed moments results from Refs.~\cite{Falcioni:2023luc,Falcioni:2025hfz}.
Moreover, we confirm the prediction from ref.~\cite{Falcioni:2025hfz} for the coefficient of a particular weight-five function in Mellin space, defined by eq.~(5) therein.
In the rational part of our all-$n$ result, a denominator factor $(n-2)$ is present, which corresponds to a spurious divergence at $n=2$.
As was pointed out in ref.~\cite{Falcioni:2025hfz}, an interesting consequence is that the Mellin moment $n=2$ receives additional contributions proportional to $\zeta_3$ from the rational part due to
\begin{equation}
  \frac{\frac{3}{4} + \, S_{-2}(n)}{n-2} = \frac{7}{4} - \frac{3}{2} \zeta_3 + \order{n-2} \,.
\end{equation}
We confirm that these contributions are explicitly generated from our results and in agreement with ref.~\cite{Falcioni:2025hfz}.

We then perform an inverse Mellin transformation as implemented in the \texttt{HarmonicSums} package~\cite{Ablinger:2009ovq,Ablinger:2012ufz} to the corresponding analytic four-loop pure-singlet splitting function.
The results contain harmonic polylogarithms~\cite{Remiddi:1999ew,Gehrmann:2001pz} and products of $\zeta$ values up to weight six.
For numerical evaluations and expansions of the harmonic polylogarithms, we employ the package \texttt{HPL} \cite{Maitre:2005uu} and the implementation \cite{Vollinga:2004sn} of $G$ functions in \texttt{GiNaC}.

\begin{figure*}[ht]
  \centering
  \includegraphics[width=0.6\linewidth]{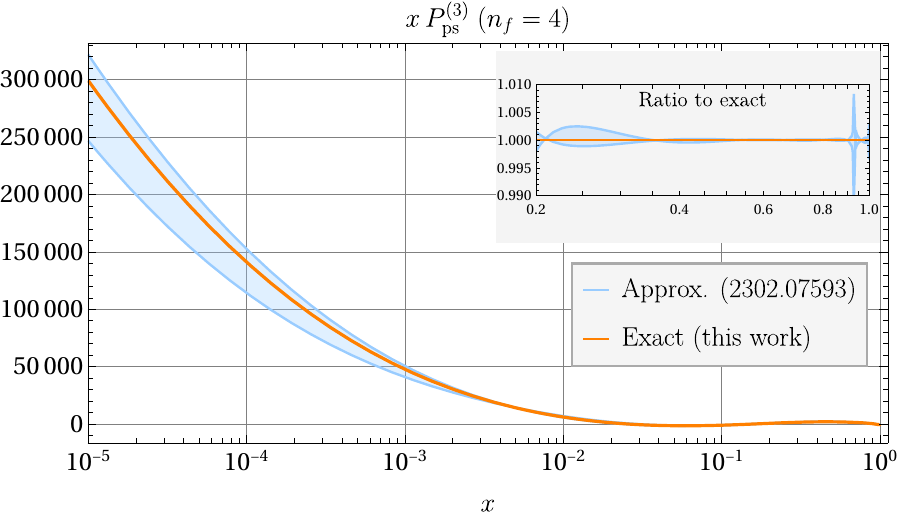} ~
  \caption{
    The four-loop pure-singlet splitting function $P^{(3)}_\mathrm{ps}$, evaluated for $\nf=4$ and compared to approximations from ref.~\cite{Falcioni:2023luc}. The function is multiplied by $x$ for presentation.
  }
  \label{fig:P3results}
\end{figure*}

With the exact analytic results at hand, we can derive its behavior in the small $x$ limit. We obtain
\begin{widetext}
  \begin{align}
    \nonumber
    \frac{P_{\mathrm{ps}}^{(3)}}{\nf}  \xrightarrow{x\to 0} 
      & \frac{1}{x} \; \left(\log^2(x) \; \left[\frac{10496}{243} C_A^2 C_F + \zeta_3 \, \frac{256}{3} C_A^2 C_F \right]  + \log(x)\; m^{(3)}_{-1,1}  + m^{(3)}_{-1,0} \right)\\ \nonumber
      & + \log ^6(x) \; \left[\frac{4}{9} \nf C_F^2-\frac{2}{9} C_F \left(-4 C_A C_F+4 C_A^2+3 C_F^2\right)\right]\\\nonumber
      & + \log ^5(x) \; \left[\frac{4}{45} C_F \left(-107 C_A C_F+68 C_A^2+32 C_F^2\right)+\frac{32}{15} \nf C_F^2\right] \\\nonumber
      & + \log ^4(x) \; \Big[-\frac{4}{27} \nf^2 C_F -\frac{4}{27} \nf C_F \left(81 C_A-476 C_F\right) - \frac{1}{27} C_F \left(-541 C_A C_F+1709 C_A^2+1557 C_F^2\right) \\
      &   \qquad +  \zeta_2 \, \left(-\frac{136}{3} C_A C_F^2+\frac{34}{9} C_A^2 C_F+\frac{524}{9} C_F^3+64 \frac{(d_F^{abcd})^2}{N_c}\right)\Big] + \ldots{}
    \label{eq:smallx}
    \intertext{with the previously unknown coefficients of the subleading logarithms at leading power} 
    \nonumber
    m^{(3)}_{-1,1} =  
      & \frac{2015440}{729} C_A^2 C_F - \frac{13688}{27} C_A C_F^2 + \frac{92800}{729} \nf C_A C_F - \frac{210176}{729} \nf C_F^2 \\\nonumber
      & \quad + \zeta_2 \, \left(-\frac{4576}{9} C_A^2 C_F -\frac{2944}{81} \nf C_A C_F +\frac{5888}{81} \nf C_F^2\right) \\\nonumber
      & \quad + \zeta_3 \, \left(-\frac{1024}{9} C_A^2 C_F +\frac{2432}{9} C_A C_F^2 +\frac{128}{3} \nf C_A C_F -\frac{256}{9} \nf C_F^2\right) \\ 
      & \quad + \zeta_2^2 \, \left(-\frac{3008}{45} C_A^2 C_F +\frac{256}{5} C_A C_F^2\right)  \,, 
    \intertext{and}
    \nonumber
    m^{(3)}_{-1,0} = 
      & \frac{45387703}{2187} C_A^2 C_F - \frac{842548}{243} C_A C_F^2 + \frac{34184}{27} C_F^3 + \nf \left(\frac{1453406}{2187} C_A C_F - \frac{4127560}{2187} C_F^2\right) + \frac{64}{27} \nf^2 C_F \\\nonumber
      & \quad + \zeta_2 \, \Big(-\frac{796624}{243} C_A^2 C_F + \frac{78736}{243} C_A C_F^2 - \frac{2480}{3} C_F^3 + \nf \left(-\frac{1928}{9} C_A C_F + \frac{2944}{9} C_F^2\right)\Big) \\\nonumber
      & \quad + \zeta_3 \, \Big(-\frac{26512}{9} C_A^2 C_F + \frac{57520}{81} C_A C_F^2 - \frac{7040}{27} C_F^3 + \nf \left(-\frac{320}{9} C_A C_F + \frac{1792}{9} C_F^2\right) - \frac{128}{9} \nf^2 C_F\Big) \\\nonumber
      & \quad + \zeta_2^2 \, \Big(-\frac{65984}{45} C_A^2 C_F + \frac{2912}{9} C_A C_F^2 + \frac{8128}{45} C_F^3 + \nf \left(\frac{8768}{135} C_A C_F - \frac{512}{15} C_F^2\right)\Big) \\\nonumber
      & \quad + \zeta_5 \, \left(-\frac{17504}{9} C_A^2 C_F + \frac{12544}{9} C_A C_F^2 - \frac{1792}{3} C_F^3\right) \\
      & \quad + \zeta_2 \zeta_3 \, \left(\frac{512}{3} C_A^2 C_F - \frac{4480}{9} C_A C_F^2 + \frac{3584}{9} C_F^3 - 2048 \frac{(d_F^{abcd})^2}{N_c}\right).
  \end{align}
\end{widetext}
The remaining terms shown in \cref{eq:smallx} agree with the prediction of ref.~\cite{Davies:2022ofz}, except for the contribution proportional to $\zeta_2$ at $\log^4(x)$.
The prediction of ref.~\cite{Davies:2022ofz} is
\begin{align*}
  \eval{\frac{P_{\mathrm{ps}}^{(3)}}{\nf}}_{\begin{array}{l} x\to 0\\ \zeta_2 \log^4(x) \end{array}} = -\frac{148}{3} C_A C_F^2+\frac{64}{9} C_A^2 C_F+\frac{524}{9} C_F^3 \;, 
\end{align*}
which is missing the $(d_F^{abcd})^2$ color structure. Similarly to what was observed in \cite{Gehrmann:2026qbl}, it does conspire to agree with our result in the leading-color approximation,
requiring a deeper structural explanation \cite{Moch:2026qsw}.

To quantify the effect of obtaining the exact analytic results on the PDF evolution, we compare to the approximation from \cite{Falcioni:2023luc} in \cref{fig:P3results}.
We observe that the approximation is in excellent agreement within its uncertainties with the exact result throughout the considered $x$ range from $10^{-5}$ to $1$. Let us note that this is consistent with the observation from \cite{Gehrmann:2023cqm} where it was found that the $\nf^2$ contribution to $P^{(3)}_\ps$ alone 
does not agree well with the similarly constructed approximation at small $x$. The reason is that the coefficient of the leading divergence $\frac{\log(x)}{x}$ of that contribution is 
next-to-leading in the full $P^{(3)}_\ps$ and was not known exactly.
For the complete $P^{(3)}_\ps$ the leading term is of higher logarithmic order  $\frac{\log^2(x)}{x}$ and
is known exactly \cite{Catani:1994sq}, such that the disagreement in the next-to-leading logarithmic contributions happens to be subdominant.

The all-$n$ result for $\gamma^{(3)}_\ps$, the splitting function $P^{(3)}_\ps$, and its small-$x$ expansion through next-to-leading power are provided in ancillary files.
To facilitate their implementation in parton evolution, we also provide a precise numerical fit to the remainder after subtracting this expansion from $P^{(3)}_\ps$, using a polynomial ansatz in $x$, $\log(x)$, and $\log(1-x)$. The fit achieves absolute and relative deviations from the exact result of at most $10^{-5}$ and $10^{-8}$, respectively.

\section{Summary and Outlook}

We have determined the four-loop pure-singlet splitting function analytically for arbitrary Mellin moments.
Combined with the non-singlet result~\cite{Gehrmann:2026qbl}, it completes the quark-to-quark splitting function at this order and removes the uncertainty associated with its previous approximations.

Our result agrees with all available partial results.
As in the non-singlet case~\cite{Gehrmann:2026qbl}, we find a discrepancy with the small-$x$ resummation prediction~\cite{Davies:2022ofz} in the subleading-color contribution proportional to $\zeta_2$ at next-to-next-to-leading logarithmic level.
Understanding the structural origin of this discrepancy remains an open question.

The exact $\gamma_{qq}^{(3)}$ also supplies the quark contribution to the singlet trace $\gamma_{qq}+\gamma_{gg}$, whose reciprocity structure can help reconstruct $\gamma_{gg}^{(3)}$ from fixed Mellin moments~\cite{Chen:2020uvt,Kniehl:2026eij}.

\begin{acknowledgments}
This work has been supported by the European Research Council (ERC) 
under the European Union's Horizon 2020 research and innovation programme grant agreement 101019620 (ERC Advanced Grant TOPUP).
\end{acknowledgments}

\newpage

\bibliography{main}

\end{document}